\documentstyle[preprint,aps]{revtex}
\begin{document}
\draft
\title{RELIC GRAVITATIONAL WAVES AND LIMITS ON INFLATION}
\author{L.P.Grishchuk}
\address{McDonnell Center for the Space Sciences, Physics Department}
\address{Washington University, St. Louis, MO 63130}
\address{and}
\address{Sternberg Astronomical Institute, Moscow University}
\address{119899 Moscow, V-234, Russia}
\maketitle
\begin{abstract}
It is shown that only a narrow class of inflationary models can
possibly agree with the available observational data on the
anisotropy of the cosmic microwave background radiation (CMBR).
These models may be governed by ``matter'' with the effective
equation of state $-1.2 <p/ \epsilon < -0.6$ which includes the
De-Sitter case $p/ \epsilon = -1$.
\end{abstract}
\vskip 2cm
\pacs{PACS numbers: 98.80.Cq, 98.70.Vc, 04.30.+}
\newpage
The recent discovery of the angular variations in the CMBR~[1] has
strongly sharpened the issue of the origin and nature of the
long-wavelength cosmological perturbations.  The consequences of
inflationary models are under active investigation (see, for
instance, a recent paper~[2] and references therein).  It seems
to me that only a narrow class of the inflationary models,
discussed below, can possibly avoid theoretical and observational
inconsistencies.

I am considering here relic gravitational waves.  The variable
gravitational field of all cosmological models (unless the
cosmological scale factor $a(\eta )$ is such that
$a^{\prime\prime} =0$), and inflationary models as a particular
case, inevitably generate gravitational waves~[3].  (The graviton
creation in FRW fields was denied in the past and anisotropic
models were claimed to be the only way to get a nonzero result,
but now they are not considered anymore as a necesary
condition.)  The generating mechanism is quantum-mechanical in
its nature, and the generated perturbations are always placed in
squeezed vacuum quantum states (see~[4] and references therein).
This means that different modes of the created field are not
totally independent, as is often assumed in the literature on inflation,
but, on the contrary, some of them are highly correlated which
leads to the picture of standing waves and modulated spectra.
The generated gravitational waves inescapably produce the angular
anisotropy in the CMBR.  The angular correlation function for
$\delta T/T$ variations caused by squeezed gravitational waves
has been derived recently~[4].  We will use it here in our
analysis.

The dimensionless gravity-wave field, with all the normalization
factors taken into account, can be written as
\begin{equation}
       h_{ij}(\eta ,{\bf x})
     = 4\sqrt{\pi} {l_{pl} \over a(\eta )}
       {1\over (2\pi )^{3/2}} \int_{-\infty}^\infty d^3{\bf n}
       \sum_{s=1}^2 p_{ij}^s({\bf n}) {1\over \sqrt{2n}}
\left[ c_{\bf n}^s (\eta ) e^{i{\bf nx}}
       + c_{\bf n}^{s^{\dag}} (\eta ) e^{-i{\bf nx}} \right]
\end{equation}
where the scale factor
$a(\eta )$, $ds^2=a^2(\eta )(d\eta^2 - dx^2 -dy^2 -dz^2)$,
has the dimension of length and $l_{pl}=(G\hbar /c^3)^{1/2}$
is the Planck's length, all other quantities are dimensionless.
The two ``transverse-traceless'' polarization tensors
$p_{ij}^s({\bf n})$ $(s = 1,\, 2)$ satisfy the conditions
$p_{ij}^s({\bf n})p^{s^{\prime}ij}({\bf n})=2\delta_{{ss}^{\prime}},$
$p_{ij}^s(-{\bf n})=p_{ij}^s({\bf n})$.
The time-dependent annihilation and creation operators
$c_{\bf n}^s(\eta )$, $c_{\bf n}^{{s}^{\dag}}(\eta )$
can be written (for each $s$) as
\begin{equation}
    c_{\bf n} (\eta ) = u_n(\eta )
    c_{\bf n} (0) + v_n(\eta ) c_{-{\bf n}}^{\dag} (0), \qquad
    c_{\bf n}^{\dag} (\eta ) = u_n^{\ast} (\eta )
    c_{\bf n}^{\dag} (0) + v_n^{\ast} (\eta ) c_{-{\bf n}}(0),
\end{equation}
where
$c_{\bf n} (0)$, $c_{\bf n}^{\dag}(0)$,
are the initial values of the operators taken at some $\eta =\eta_0$
long before the interaction became effective. The complex functions
$u_n(\eta )$, $v_n(\eta )$ satisfy the equations
\begin{equation}
  iu_n^\prime = nu_n +i(a^\prime /a)v_n^\ast , \qquad
  iv_n^\prime = nv_n +i(a^\prime /a)u_n^\ast
\end{equation}
where ${}^\prime =d/d\eta$, $|u_n|^2 - |v_n|^2 =1$ and $u_n(0)=1$, $v_n(0)=0$.
It follows from these equations that the function
$\mu_n(\eta)\equiv u_n(\eta )+ v_n^{\ast}(\eta )$
obeys the equation
$\mu_n^{\prime\prime} +(n^2-a^{\prime\prime}/a)\mu_n=0$
which is precisely the equation for classical complex
$\mu$-amplitude~[3] of the gravity-wave field.

The (Bogoliubov) transformation (2) can be written in a form
involving the two-mode squeeze operator which demonstrates
the inevitable appearance of squeezing in this kind of
problems. In the Schr\"odinger picture, the initial vacuum state
$|0\rangle$,  $c_{\bf n}(0)|0\rangle =0$,
evolves into a strongly squeezed vacuum state.

The angular correlation function for $\delta /T$
(see Eqs.~(12), (13) in Ref.~[4]) can be rearranged
by using the ``summation theorem''~[5] and the formulas relating the
Gegenbauer polynomials to the associated Legendre polynomials~[6]
and cast into an elegant exact form which directly
involves the Legendre polynomials $P_l(\cos\delta )$:
\begin{equation}
   \langle 0| {\delta T \over T}
   (e_1^k) {\delta T \over T} (e_2^k)|0 \rangle
 = l_{pl}^2 \sum_{l=2}^\infty K_l P_l(\cos\delta )
\end{equation}
where $K_l = (2l+1)l(l+1)[l(l+1)-2]F_l$,
\begin{equation}
   F_l = \int_0^\infty n^2 \left| \int_0^{w_1}
   {J_{l+{1\over 2}} (nw) \over (nw)^{5/2}}
   f_n(\eta_R-w)dw \right|^2 dn
\end{equation}
and
\[
f_n(\eta_R-w) = {1\over\sqrt{2n}} (\mu_n /a)^\prime \, .
\]
The two unit vectors $e_1^k$, $e_2^k$ point out in the directions of
observation and $\delta$  is the angle between them.
The photons of CMBR are assumed to be emitted at
$\eta =\eta_E$ and received at (present) time
$\eta=\eta_R$; $w=\eta_R -\eta$, $w_1=\eta_R -\eta_E$
(the Sachs-Wolfe effect~[7]).
Note that the correlation function, with no additional assumptions made
whatsoever, is rotationally symmetric and its multipole expansion begins
from $l = 2$.

The derived formula is universal and can be used with
arbitrary $a(\eta )$. We will apply this formula to simple models
consisting of three consecutive stages of expansion: inflationary
($i$-stage), radiation-dominated ($e$-stage) and matter-dominated
($m$-stage)~[8].

The scale factor of the model can be written at the three
stages as follows.\\
$i$-stage:
\[
   a(\eta )= l_0 | \eta |^{1+\beta}, \quad
   \eta \leq \eta_1 \, , \quad \eta_1 <0 ,
\]
$e$-stage:
\[
    a(\eta )= l_0 a_e (\eta - \eta_e) , \quad
    \eta_1 \leq \eta \leq \eta_2
\]
where $a_e =-(1+\beta )| \eta_1 |^\beta$,
$\eta_e ={\beta\over 1+\beta}\eta_1$,\\
$m$-stage:
\[
   a(\eta )= l_0 a_m (\eta - \eta_m )^2, \quad \eta_2 \leq \eta ,
\]
where $a_m =[a_e /4(\eta_2-\eta_e)]$,
$\eta_m = -\eta_2 +2\eta_e$.
The functions $a(\eta )$, $a^\prime (\eta )$ are continuous at
$\eta = \eta_1$ and $\eta =\eta_2$.  All expanding models with
$1+\beta < 0$ ($\eta$ must be negative if $1+\beta < 0$) are
inflationary in the sense that the length scale equal to the
Hubble radius at some early time of expansion can grow at all
three stages up to, at least, the size of the present day Hubble
radius $l_H =a^2/a^\prime$, $\eta = \eta_R$.  The case
$\beta =-2$ corresponds to the De-Sitter expansion, the cases
$\beta < -2$ correspond to the so called power-law inflation
$(a(t)\sim t^m$, $m>1$) and the cases $-2<\beta <-1$ (apparently,
not having been analyzed before) correspond to the law of
expansion $a(t) \sim |t|^m$, $m<-1$, $t<0$.  The $i$-stage is
governed by ``matter'' with the effective equation of state
$p=q(\beta )\epsilon$, where $q(\beta )=(1-\beta )/3(1+\beta )$
and $q(\beta )$ varies from $-1/3$ to $-\infty$ for
$-\infty < \beta <-1$.  Expansion is accompanied by the growth of
energy density and curvature if $-2<\beta <-1$.

In realistic cosmological models
$a(\eta_E)/a(\eta_R)\approx 10^{-3}$,
$a(\eta_2)/a(\eta_R)\approx 10^{-4}$.  Also, one has
$3.10^{-32}<a(\eta_1)/a(\eta_R) < 3.10^{-12}$ if one wants to
commence the $e$-stage at densities not lower than the nuclear
and not higher than the Planckian, or, in other words, if one
wants the Hubble radius $l_i$ at the end of inflation,
$l_i =-l_o(1+\beta )^{-1} |\eta_1|^{2+\beta}$,
to be in the interval $1<l_i/l_{pl}<10^{40}$.

To define the numerical values of $\eta$ it is convenient to choose
$\eta_R-\eta_m=1$.  Then,
$|\eta_1|\approx 5^{-2/ \beta} |1+\beta|^{-1/ \beta}$
$(l_o/l_H)^{-1/ \beta}$ and
$(l_{pl}/l_o)^2 \approx (25l_{pl}/l_H)^{2+\beta}$
$|1+\beta |^{2(1+\beta )} (l_i/l_{pl})^\beta$.
The wavelength
$\lambda =2\pi a/n$ equal to $l_H$ has the wave number $n_H=4\pi$,
and the wavelength equal to the Hubble radius at
$\eta=\eta_2$ corresponds to
$n_m\approx 4\pi\cdot 10^2$.
The minimally sufficient inflation should begin not later than at
$\eta_b =(1+\beta )/2$, its variable gravitational field
generates waves with wavelengths up to
$l_H$.  Inflation that started earlier inevitably generates the longer
waves also.

The gravity-wave eguation using the scale factor of the form
$a(\eta )= a_o\eta^{1+\beta}$ has been solved and the relation
between the initial and final amplitudes has been derived earlier~[3].
The waves start oscillating with the amplitude
$B(n)$ which is related to the initial amplitude $A(n)$ by
$B^2(n)\sim A^2(n)(n\eta_1)^{2(1+\beta )}$
(one can use formula (5b) from the second Ref.~[3] in which
the interpretation of the participating amplitudes should
be reversed because, for $1+\beta <0$,
the condition $(n\eta )^2 \gg 1$ is
satisfied initially and gets violated later on).
Since the initial (vacuum) spectrum goes as
$A(n)\sim n$ this leads to $B(n)\sim n^{\beta +2}$ and
$B(n)\sim n^o$ for $\beta =-2$, that is, in case
$\beta =-2$, all waves start oscillating with the same amplitude
(the Harrison-Zeldovich spectrum). In case of
$\beta < -2$ the spectrum gets ``tilted'' by increasing
the relative contribution of longer waves, and in case of
$\beta > -2$ --- by increasing the relative contribution of shorter waves.

The exact solution to Eq.~(3) for the complex function $\mu_n(\eta )$
satisfying the required initial data and continuous with its
first time-derivative at the joining points
$\eta_1, \, \eta_2$ has the following form.\\
$i$-stage:
\[
   \mu_n(\eta ) = (n\eta )^{1/2} [A_1 J_{\beta+1/2} (n\eta )
                + A_2 J_{-(\beta +1/2)} (n\eta )]
\]
(for a technical simplification we work solely with the Bessel
functions and exclude the half-integer $\beta$'s but the final
result will be free of this limitation) where
\[
   A_1 = -{i\over \cos\beta\pi} \sqrt{{\pi\over 2}} \,
         e^{i(x_0+\pi\beta /2)},  \quad
   A_2 = iA_1e^{-i\pi\beta} \quad {\rm and} \quad
   x_o \equiv n\eta_o$, \quad $n|\eta _o| \gg 2\pi |1+\beta |
\]
$e$-stage:
\[
  \mu_n(\eta ) = B_1\, e^{-iny} + B_2 \, e^{iny}
\]
where $y\equiv n(\eta - \eta_e )$,\\
$m$-stage:
\begin{equation}
   \mu_n(\eta ) = \sqrt{{\pi z \over 2}}
   (C_1 J_{3/2}(z)+C_2J_{-3/2}(z)) \, ,
\end{equation}
\begin{equation}
   \left( {\mu_n \over a} \right)^\prime
 = -{n\over a} \sqrt{{\pi z \over 2}}
    (C_1 J_{5/2}(z) - C_2J_{-5/2}(z))
\end{equation}
where $z\equiv n(\eta -\eta_m)$. Note that
$J_{-3/2}(z)$ represents the so-called decaying solution which
is necessarily present.

The coefficients $B_1$, $B_2$, $C_1$, $C_2$,
are determined by the continuos joining of the solutions. In particular,
\[
  -C_1 = B_1(\alpha_2 +\beta_2^\ast )+ B_2(\alpha_2^\ast+\beta_2)\, , \quad
 -iC_2 = B_1(\alpha_2 -\beta_2^\ast )- B_2(\alpha_2^\ast-\beta_2)
\]
where
$\alpha_2 = e^{iy_2} (8y_2^2 -1+i4y_2)/8y_2^2$,
$\beta_2=-e^{i3y_2}/8y_2^2$,  $y_2\equiv n(\eta_2-\eta_e )$.
We are only interested in modes that have interacted with the barrier
$U(\eta )=a^{\prime\prime}/a$ and have been generated quantum-mechanically.
Their wave numbers obey the condition
$n|\eta_1| \ll 2\pi |1+\beta|$ and for them
\[
   B_1\approx -B_2 \approx {1\over 2} e^{ix_o}
     (\beta+1)\psi(\beta)(n\eta_1 )^\beta \equiv B \, ,
\]
where
\[
 \psi (\beta ) \equiv \sqrt{{\pi\over 2}} e^{i\pi\beta /2}
\left[ \cos \beta\pi \, 2^{\beta+1/2}\,
       \Gamma(\beta+3/2)\right]^{-1} \, , \quad
       |\psi (\beta )| = 1 \quad {\rm for}\quad \beta=-2 \, .
\]
The values of $C_1$, $C_2$ depend on whether $y_2 \gg 1$ or $y_2 \ll 1$.
For relatively short waves, $n\gg n_m$, one has approximately
$C_1 \approx -2iB\sin y_2$, $C_2\approx 2iB\cos y_2$.
For longer waves, $n\ll n_m$, one has
$C_1\approx -{3i\over 2}By_2^{-1}$,
$C_2\approx -{8i\over 45}By_2^4$, $C_2\ll C_1$.
The additional large factor $y_2^{-1}$ in $C_1$ reflects additional
amplification of waves at $m$-stage.

For a qualitative description of amplitudes and spectral slopes we
introduce ``characteristic'' spectral components of the field:
$h(n)=l_{pl} n|\mu_n| /a$, and its first time-derivative:
$h^\prime (n)=l_{pl}(|\mu_n| /a)^\prime$.  We have
$h(n)\sim {l_{pl}\over a}n^{\beta +1}\sin [n(\eta -\eta_e )]$,
$h^\prime (n)\sim {l_{pl}\over a}n^{\beta +1}\cos [n(\eta -\eta_e )]$
for $n\gg n_m$, and
$h(n)\sim {l_{pl}\over a}n^\beta \cos [n(\eta -\eta_m )]$,
$h^\prime (n) \sim {l_{pl} \over a} n^\beta \sin [n(\eta - \eta_m)]$
for $n_H \gg n \gg n_m$.  For $n\ll n_H$ one can use the
approximation $z\ll 1$ in Eqs.~(6), (7) and obtain\\
$h(n)\approx {l_{pl}\over l_o}|\psi (\beta )|n^{\beta+2}$,
$h^\prime (n)\approx {l_{pl}\over 5l_o}|\psi (\beta )|zn^{\beta +2}$.

We will now start deriving restrictions on inflationary models.
Astrophysically interesting and consistent values of $h$ require
$h(n_H)$ to be not much larger or much smaller than $10^{-4}$.
The mean square value of the field diverges in the limit of small
$n$ ({\it i.e.} wavelengths much longer than $l_H$) for all
$\beta \le -2:$
$\langle h^2\rangle\approx (l_{pl}/l_o)^2
|\psi (\beta )|^2\int {dn\over n}n^{\beta +2}$.
This does not allow the duration of inflation to be arbitrarily long.
In case of $\beta =-2$ the divergence is logarithmic and the restriction
on duration is very mild but it becomes increasingly severer for
$\beta < -2$.  For $\beta\approx -7$, even the minimally sufficient
duration of inflation does not help as the $h(n_H)$ strongly exceeds
$10^{-4}$ even if the largest allowed $l_i$,
$l_i\approx 10^{40}\, l_{pl}$,
is chosen. On the other  hand, for $\beta >-2$
there is no problem with the long-wavelength divergence
but the amplitudes very quickly become too small. For
$\beta \approx -1.8$, the $h(n_H)$
becomes smaller than the required level even if the smallest allowed
$l_i$, $l_i \approx l_{pl}$, is chosen. (One may argue that the values
$l_i \ll l_{pl}$ do not necesserily invalidate the analysis and
can also be allowed, although they imply the over-Planckian densities,
since the wavelengths of our interest are much longer than $l_{pl}$
all the way up from the beginning of inflation.  However, we will confine
in this paper to the requirement $l_i > l_{pl}$ which leads
to $\beta < -1.8$.)

Additional restrictions come from the $\delta T/T$ considerations.
We will be interested in the lower multipoles
$l$, $l \le 30$, to which the waves $n \gg n_m$ give a negligibly
small contribution. Neglecting the second term in Eq.~(7) one
can write Eq.~(5) as
\[
  F_l \approx {9\pi \over 4l_o^2} |\psi (\beta )|^2
  \int_0^{n_m} n^{2\beta+3} \left| \int_o^{nw_1}
  {J_{l+1/2} (x)\over x^{5/2}}
  {J_{5/2} (n-x)\over (n-x)^{3/2}} dx \right|^2 dn
\]
where $w_1 \approx 1-3.10^{-2}$.
In the limit of long waves one can use the small
argument approximation for the Bessel functions and write
\[
   F_l \approx {9\pi\over 4l_o^2} |\psi (\beta )|^2
   \phi^2(l) \int n^{2\beta+2l+3} dn \, ,
\]
where
$\phi (l) = [15\sqrt\pi 2^l \Gamma (l+1/2)l(l-1)]^{-1}$.
The quadrupole component $K_2$ is divergent in the limit of small
$n$ for all $\beta \le -4$ so these values of $\beta$
should be excluded unless the duration of inflation is precisely
tuned. (Interestingly enough, the minimally sufficient inflation may be a
likely outcome of certain quantum-cosmological models~[9].)

The multipole distributions $K_l$ for some models
have been numerically computed by A. Wiseman.
For each $l$, the main contribution to $F_l$ comes,
as one might expect, from the interval
of $n$ between $n\approx l$ and $n\approx 2l$.
In particular, $K_2$ is dominated by waves longer than $l_H$~[10].
The absolute values of $K_l$ depend on $\beta$ and $l_o$.  For
instance, the quadrupole component $10^2 (l_o /l_{pl})^2\, K_2$
has the following numerical values:  9.7, 7.3, 5.6, 7.4, if one
takes $\beta =-1.8$, $-2.0$, $-2.4$, $-3.0$, correspondingly.
The multipole distributions normalized to $K_2$, that is the functions
$K_l/K_2$, are independent of $l_o$.  They are shown in Fig.~1.
For completeness we have included also the extreme cases
$\beta =-1.6$, $-1.1$, though they would imply strongly over-Planckian
densities at the end of inflation, as was explained above.
It is worth noting that the quadrupole component exceeds other multipoles
and $K_3/K_2 < 0.6$ for all models with $\beta < -1.8$.
The results for the $\beta =-2$ case are in qualitative agreement
with those in Ref.~11.  It remains to be seen
which of these distributions can survive after comparison
with the detailed COBE-type observations. One should bear in mind,
of course, that predictions for $K_l$ are statistical and should be
augmented with variancies based on the statistics of squeezed quantum
states.

In conclusion, the inflationary models governed by ``matter'' with
the effective equation of state $-1.2 < p/ \epsilon < -0.6$
seemingly avoid theoretical difficulties and some of them can possibly
withstand comparison with the observations. (The models include
the De-Sitter case if the duration of expansion was not
excessively long.) Taking into account the
density perturbations can only make this interval narrower.
\newpage

I would like to thank Alan Wiseman for his help.

This work was supported in part by NASA grant NAGW 2902 and NSF
grant 89-22140.

\newpage
\begin{figure}
\caption{The normalized multipole distributions.}
\end{figure}
\end{document}